\documentclass[11pt,tightenlines,eqsecnum,floats,aps,nofootinbib,prd,showpacs]{revtex4}
\usepackage{hyperref}
\usepackage{amsmath,amssymb,amsfonts,amsthm,amscd}
\usepackage{graphicx}
\usepackage{enumerate}
\usepackage{colordvi}
\usepackage{units}
\usepackage{epsfig}
\usepackage{natbib}
\usepackage{enumerate}
\usepackage{colordvi}
\usepackage{multirow}
\usepackage{afterpage}
\usepackage{pdflscape}
\usepackage{subcaption}
\def\be{\begin{equation}}
\def\ee{\end{equation}}
\def\ba{\begin{eqnarray}}
\def\ea{\end{eqnarray}}

\def\dif{d}

\begin{document}

\title{Measures in Multifield Inflation}

\author{Richard Grumitt}
\email{richard.grumitt@st-hildas.ox.ac.uk}
\affiliation{Beecroft Institute of Particle Astrophysics and Cosmology, Department of Physics,
University of Oxford, Denys Wilkinson Building, 1 Keble Road, Oxford, OX1 3RH, UK}

\author{David Sloan}
\email{david.sloan@physics.ox.ac.uk}
\affiliation{Beecroft Institute of Particle Astrophysics and Cosmology, Department of Physics,
University of Oxford, Denys Wilkinson Building, 1 Keble Road, Oxford, OX1 3RH, UK}

\begin{abstract}

We examine the classical dynamics of multifield inflation models with quadratic potentials. Such models are shown to have inflationary attractors in phase space, consistent with the stretching of phase space trajectories along the volume factor of the universe during inflation. Using the symplectic structure associated with Hamiltonian systems we form a measure on the phase space, as initially proposed by Gibbons, Hawking and Stewart. This is used to calculate lower bounds on the probabilities of observational agreement (i.e. the probability the model gives a value for the spectral index within the region $n_{s}=0.968\pm{0.006}$) for equal mass two and three field models with quadratic potentials, giving values of 0.982 and 0.997 respectively. We derive the measure for a general $N$-field model and argue that as the number of fields approaches infinity, the probability of observational agreement approaches one. 

\end{abstract}

\pacs{04.20.Dw,04.60.Kz,04.60Pp,98.80Qc,04.20Fy}
\maketitle

\section{Introduction}

\subsection{Inflation}

Since it was originally developed by Guth, Linde, Albrecht and Steinhardt, inflationary theory has become a pillar of modern cosmology. The theory was initially motivated by the horizon, flatness and magnetic monopole problems~\cite{GuthInflation,Linde,Albrecht}. Soon after it was proposed it was realised that inflation could also produce the perturbations that would act as the seeds of structure formation~\cite{Starobinsky,Pi,Hawkpert,Bardeen,Kahn,Pi2,Dodelson}. We will work throughout in Planck units ($G=c=\hbar=1$).

Let us first consider how we can implement inflation in the early universe. We begin by asking what generic properties the field driving inflation will have. From the Friedmann equation:
\begin{equation}
H^2\equiv\left(\frac{\dot{a}}{a}\right)^2=\frac{8\pi}{3}\rho-\frac{k}{a^{2}}
\end{equation}
 where $H$ is the Hubble parameter, $a$ is the scale factor, $\rho$ is the energy density of the universe and $k$ is the curvature parameter. Differentiating this equation and using the equation for energy conservation in an FRW universe ($\dot{\rho}+3\frac{\dot{a}}{a}(\rho+P)=0$, where $P$ is the pressure), we obtain the Raychaudhuri equation:
\begin{equation}
\frac{\ddot{a}}{a}=-\frac{4\pi}{3}(\rho+3P)
\end{equation}
For inflation we require that $\ddot{a}>0$ so that the expansion rate is accelerating. This is satisfied if our matter has an equation of state that satisfies $P<-\rho/3$. This cannot be achieved with normal matter or radiation however we can obtain a natural implementation with some generic scalar field/s. Consider the equation of state parameter for a single, spatially constant scalar field $\phi$~\cite{Peacock}:
\begin{equation}
w\equiv\frac{P}{\rho}=\frac{\frac{1}{2}\dot{\phi}^2-V(\phi)}{\frac{1}{2}\dot{\phi}^2+V(\phi)}
\end{equation}
 where $V(\phi)$ is the potential of our scalar field. We can see here that if $\frac{1}{2}\dot{\phi}^{2}\ll V(\phi)$ our equation of state parameter $w\approx-1$, satisfying the condition for an accelerating expansion rate. This slow roll inflation where the kinetic energy of the field is negligible is precisely the implementation most commonly used and will be the focus of our attention. This can be visualised as the inflaton (the particle associated with the scalar field) slowly rolling down the field potential. We can characterise the slow roll for a set of $N$ scalar fields $\phi_{i}$ by defining slow roll parameters as in~\cite{Bassett}:
\begin{equation}
\epsilon\equiv\frac{1}{16\pi}\frac{\sum_{i}\left(\partial V/\partial\phi_{i}\right)^{2}}{V^{2}}
\end{equation}
\begin{equation}
\eta_{\sigma\sigma}\equiv\frac{1}{8\pi}\frac{V_{\sigma\sigma}}{V}
\end{equation}
 where for a general multifield model, $V_{\sigma\sigma}$ is the second derivative of the potential along the field direction in the field space defined by:
\begin{equation}
\sigma=\int\sum_{i}\hat{\sigma}_{i}\dot{\phi}_{i}\dif t
\end{equation}
 where $\hat{\sigma}_{i}$ is the unit vector along the field direction:
\begin{equation}
\hat{\sigma}_{i}=\frac{\dot{\phi}_{i}}{\sqrt{\sum_{k}\dot{\phi}_{k}\dot{\phi}_{k}}}
\end{equation}
Other slow roll parameters can be defined but these are the only parameters that will be relevant. During the slow roll phase $w\approx-1$.

It is important to note here that there is currently no verified particle physics mechanism for inflation~\cite{Dodelson}. Given this our aim will not be to ground the discussion of inflationary mechanisms in any known physics, but rather to examine the possibilities presented by multiple scalar fields.

\subsection{The measure problem}

The measure problem in classical inflationary cosmology may be framed as follows; given some model for inflation and the set of all possible model universes, how do we go about counting fractions of these universes i.e. which measure should be used when counting over the universes? Moreover, given a suitable measure, what is the probability that some inflationary model yields a universe observationally similar to our own? The precise meaning of observationally similar in this context will be made clear in section 3.

Inflation was developed to provide an explanation for why the universe did not need to exist in a very peculiar, finely tuned initial state. However if we were to find that inflation was incredibly unlikely to produce a universe like our own, we would be hard pressed to claim it as a complete solution to such tuning problems. We would effectively be replacing a special initial state with another special initial state. Moreover, whilst the work discussed here focuses on inflationary models and their measures, the study of measures has potential applications for more general dynamical problems, allowing us to calculate the probabilities of systems exhibiting certain properties \cite{SloanMin}.

Significant work has been carried out in trying to answer the above questions. Gibbons, Hawking and Stewart proposed the Liouville measure as a natural measure on the set of model universes. This is formed from the symplectic structure associated with the Hamiltonian phase space of an FRW universe coupled to  matter. A Hamiltonian system possesses the canonical coordinates $(q_{i},p_{i})$, where $q_{i}$ are the generalised positions for our system and $p_{i}$ the associated generalised momenta. For a system with $n$ degrees of freedom we therefore have a $2n$ dimensional phase space equipped with a symplectic structure given by:
\begin{equation}
\omega=\sum_{i=1}^n \dif q_{i}\wedge\dif p_{i}
\end{equation}
 where the $\wedge$ is the antisymmetric wedge product. The symplectic structure is a differential two-form that is a property of the Hamiltonian phase space. In general a $p$-form is a completely antisymmetric, rank $(0,p)$ tensor i.e. it acts as a map from a collection of $p$ vectors to $\mathbb{R}$ (the set of real numbers). In the language of index notation, a $p$-form is a completely antisymmetric tensor with $p$ covariant indices. Given the antisymmetry of the wedge product we can see immediately that $\dif f\wedge\dif z=-\dif z \wedge\dif f$ and hence $\dif f\wedge\dif f=0$ for some $f$ and $z$~\cite{Carroll}.

The important point here is that we can form a volume element $\Omega_{L}$ with the symplectic structure by raising it to the power $n$~\cite{Turok}:
\begin{equation}
\Omega_{L}=\frac{(-1)^{n(n-1)/2}}{n!}\omega^{n}
\end{equation}
This is the Liouville volume element of our phase space which is unique up to a functional freedom i.e. this is the only $2n$-form on our phase space up to a choice of multiplicative function. If we differentiate the symplectic structure with respect to time and apply Hamilton's equations of motion to the result we find that $\dot{\omega}=0$ under Hamiltionian flow. An immediate consequence of this is that the Liouville volume element is conserved, and that phase space volume elements are conserved for a Hamiltonian system. This is simply a statement of Liouville's theorem.

Let us consider what we mean when we claim that the measure formed from the symplectic structure is the natural measure to use when counting universes. To a certain extent the phase space measure we choose is ad hoc~\cite{Schiffrin, Sloan}. If we wished we could take any function of our phase space variables to construct a new measure. However we can argue that the symplectic measure should be viewed as a natural or privileged measure on phase space. Here we appeal to Laplace's principle of indifference which tells us that we should adopt the distribution with the least amount of information content. The Hamiltonian phase space is naturally equipped with a symplectic structure and so we should use a uniform distribution on the symplectic measure to minimize the information content~\cite{Sloan}.  

One of the apparent contradictions that arose from the study of the measure problem in inflation was the existence of inflationary attractors, as noted in~\cite{Turok,Remmen,AshProbLQC,Corichi,Sloan}. Scalar field trajectories in phase space display a focusing on a very narrow range of field values as they evolve. This seems to sit in contradiction with Liouville's theorem that tells us phase space volumes should remain constant under Hamiltonian flow. The resolution to this problem was discussed in detail in~\cite{Corichi,Sloan}. The trajectories do indeed display the focusing on a narrow range of field values, however the key observation is that as these solutions inflate the trajectories are stretched dramatically along the volume factor axis. The total phase space volume remains conserved under evolution, it is simply stretched out along the volume factor of the universe as it expands. 

Other issues do exist in forming this measure. The most notable is the fact that, because the volume direction is non-compact, the integral over the volume factor is infinite. This must be regularised in order to yield meaningful probabilities. These issues will be discussed in detail when we consider measures and probabilities in multifield inflation.

\subsection{Measures in single field inflation}

The measure problem has received significant attention in the context of single field inflation~\cite{Hawking,Turok,Remmen,Remmen2,AshProbLQC,Corichi}. In the case of a single quadratic field the symplectic structure can be used to obtain an expression for the probability that the universe evolves to have a value for the spectral index, $n_{s}$, consistent with the Planck results \cite{Sloan}. The process of obtaining a probability from the symplectic structure will be discussed in detail when we consider multifield models. Starting the evolution from a surface of constant Hubble rate $H=1$ (reinserting units this is $H=5.72\times10^{62} $kms$^{-1}$Mpc$^{-1}$) they were able to obtain a probability of observational agreement of $P(X)=(1-10^{-5})$~\cite{Sloan}. Similarly high results were obtained earlier in the context of loop quantum cosmology in~\cite{AshProbLQC,Corichi,AshLQC}. These results are in stark contrast with an earlier result by Gibbons and Turok, who obtained a probability of there being more than 60 $e$-folds\footnote{The questions of whether we obtain more than 60 e-folds or an observationally consistent value for $n_{s}$ are similar. To have a universe with properties similar to our own we want a sufficient amount of inflation, however universes which underwent a large number of e-folds are not necessarily consistent with our own. The latter question can therefore be viewed as a refinement of the former~\cite{Sloan}.} of inflation of $~e^{-180}$~\cite{Turok}. 

The reason for the incredibly low value calculated by Gibbons and Turok was explained in~\cite{Corichi,Sloan}. Gibbons and Turok had evaluated their measure on a very late constant energy density surface (or equivalently constant $H$ surface from the Friedmann equation), assuming a uniform probability distribution of initial conditions at this time. This late in the evolution, most of the trajectories will have been focused in on a very narrow range of field values consistent with observation by the attractor. By assuming the uniform distribution at this late surface they gave an equal weighting to the trajectories focused in on the attractor to those that had not funnelled in, resulting in a low probability being calculated. Had they assumed a uniform distribution on an earlier constant $H$ surface they would have obtained a high probability for inflation.

This was extended by asking what would be required for the probabilities calculated on the two separate surfaces to be the same. It was realised that if we assume a uniform probability distribution on an early constant $H$ surface, a probability distribution is induced on later Hubble surfaces that is heavily weighted towards field values consistent with observation. This is a consequence of the attractor behaviour. Solutions are funnelled in towards a very narrow range of field values during their evolution, giving us a much higher density at these attractor regions.

\subsection{Multifield Inflation}

Most of the existing literature has focused on single field inflation. In the context of the measure problem, no major efforts have been made so far to form measures and calculate probabilities for multifield models. Indeed, the authors of~\cite{Hawking,Carroll,Turok,Remmen,Remmen2,AshProbLQC,Corichi,Sloan,AshLQC} all consider single field models when discussing measures and probabilities in inflation. Of course, single field models are simpler to study and do yield interesting behaviour but we do have good reason to study multifield inflation models. The Planck collaboration recently presented a comprehensive set of constraints on inflation driven by single field models. The results from this survey mean that single field $V(\phi)\propto\phi^2$ and natural inflation are currently disfavoured compared to models that predict a lower value of the tensor to scalar ratio $r$ (this is the ratio of amplitudes of tensor and scalar perturbations)~\cite{Planck}. It is therefore prudent to consider whether, despite these simple models being disfavoured in the single field case, they can still be viable models when driven by multiple fields of this type.

The interest in multifield inflation also extends beyond the fact that simple single field models are disfavoured by Planck data. Many current theories of beyond the standard model particle physics predict the existence of multiple scalar fields. For example, string compactifications often predict the existence of hundreds of scalar fields~\cite{Easther,Grana,Douglas,DenefDoug,Denef}. It would be surprising if only one of these fields was driving inflation. Whilst we have no experimental confirmation of any of these beyond the standard model theories, if we wish to ground inflation in them we must seriously consider the possibility of multifield inflation. 

If we are to study multifield inflation models in general, we should also wish to extend our analysis of the measure problem to these models. It was noted by Easter et al. that multifield models give a wider range of possible predictions for perturbation power spectra etc. It is therefore important to ask what sort of predictions can be considered generic~\cite{Easther}. The observational consequences of multifield models through cosmological perturbations and non-gaussianities have been studied extensively in ~\cite{Easther,Giblin, Vernizzi, Battefeld,Lyth}. Multifield models can predict large non-gaussianities in the CMB, meaning searching for non-gaussianities provides one of the main observational tests of multifield models. No significant evidence for this primordial non-gaussianity was found in the Planck 2015 results~\cite{Gauss}.

Important work on making generic predictions from multifield inflation has been carried out in~\cite{Easther}, which presented a set of generic predictions for multifield models. They found that their probability distribution functions for cosmological observables become more sharply peaked as they increase the number of fields. The same research group also presented a powerful and efficient numerical solver for multifield models, \texttt{MULTIMODECODE}, in Price et al.~\cite{Price} 

In~\cite{Easther}, Easther et al. use a simple measure, $\dif\phi_{1}\wedge\dif\dot{\phi}_{1}\wedge...\wedge\dif\phi_{N}\wedge\dif\dot{\phi}_{N}$, to count solutions in phase space. Although it is a simple measure to employ, and will give a reasonable idea of the probability distributions, it is not a well motivated measure. Following the arguments of Gibbons, Hawking and Stewart, the most natural phase space measure to use is that formed by the symplectic structure associated with the phase space. 

We will derive the equations of motion for the multiple scalar fields and implement them numerically to study their dynamics. We present derivations of the measures and expressions for the probabilities of observational agreement for simple two and three field quadratic models, along with a discussion of the issues in forming these measures. Lower bounds on the probabilities of observational agreement for these models in the case of equal mass fields are also found, with the probability increasing with the number of fields. We then derive a general $N$-field measure that can be implemented in future work and argue that the probability of observational agreement will approach 1 as the number of fields approaches infinity.

\section{The multifield equations of motion}

We wish to derive the equations of motion for $N$ scalar fields driving inflation. We start by writing down the general Einstein-Hilbert action coupled to some matter source~\cite{Carroll}:
\begin{equation}
S=\int \dif ^{4}x\sqrt{-g}(R+\mathcal{L}_{m})
\end{equation} 
 where $g$ is the metric determinant, $R$ is the Ricci scalar and $\mathcal{L}_{m}$ is the matter Lagrangian. This is true for any metric but we restrict ourselves to the flat FRW metric given by $\dif s^2=-\dif t^{2}+v^{2/3}(t)[\dif x^2+\dif y^2+\dif z^2]$, where $v(t)$ is the volume factor of the universe \footnote{This analysis can be extended to both positive and negative curvature spatial slices, however since the role of curvature is minimal during the period of inflation, we exclude this contribution for clarity of exposition.}. We adopt $v(t)$ as opposed to the more familiar scale factor $a(t)$, that is simply related by $v(t)=a^{3}(t)$. This gives a Ricci scalar:
\begin{equation}
R=2\frac{\ddot{v}}{v}-\frac{1}{3}\left(\frac{\dot{v}}{v}\right)^2
\end{equation}
For $N$ scalar fields the matter Lagrangian is simply $\mathcal{L}_{m}=\sum_{i}\frac{1}{2}\dot{\phi}_{i}^2-V(\phi_{1},...,\phi_{N})$ where $\dot{\phi}_{i}$ are the field velocities and $V(\phi_{1},...,\phi_{N})$ is the potential. This gives us a total Lagrangian of:
\begin{equation}
\mathcal{L}=v\left(R+\sum_{i=1}^N\frac{1}{2}\dot{\phi}_{i}^2-V(\phi_{1},...,\phi_{N})\right)
\end{equation}
The second derivative of the volume is a total derivative in the action so will only give boundary contributions when the action is varied. We can therefore ignore it. Taking the Legendre transform of the Lagrangian we obtain the Hamiltonian for this model universe:
\begin{equation}\label{eq:13}
\mathcal{H}=-\frac{3vH^2}{8\pi}+\sum_{i=1}^N\frac{P_{i}^2}{2v}+vV(\phi_{1},...,\phi_{N})
\end{equation} 
where $P_{i}=v\dot{\phi}_{i}$ are the canonical momenta associated with the respective fields $\phi_{i}$. In general relativity the Hamiltonian is constrained so that $\mathcal{H}=0$. The Hamiltonian constraint gives us the familiar Friedmann equation:
\begin{equation}
H^2=\frac{8\pi}{3}\left(\sum_{i=1}^N\frac{1}{2}\dot{\phi}_{i}^2+V(\phi_{1},...,\phi_{N})\right)
\end{equation}
where the Hubble parameter is given in terms of the volume factor by $H=\dot{v}/3v$. We also note that the dynamics of our system are unchanged by rescaling the volume $v\rightarrow\alpha v$ so that $(v,H,\phi_{i},P_{i})\rightarrow(\alpha v,H,\phi_{i},\alpha P_{i})$, which simply gives $\mathcal{H}$ an overall multiplicative constant factor. This point will become important when we come to calculating probabilities.

The scalar field equations of motion (i.e. the Klein Gordon equations), are obtained from Hamilton's equations of motion. For the field $\phi_{i}$ we have:

\begin{equation}
\ddot{\phi}_{i}+3H\dot{\phi}_{i}+\frac{\partial V}{\partial\phi_{i}}=0
\end{equation}

We note here that this equation has a similar form to that of a harmonic oscillator with a Hubble friction term $3H\dot{\phi}_{i}$ but with $H$ a function of the field variables. For sum-separable potentials the field coupling only arises through this same term. These are the equations of motion that form the object of our study. 

Detailed numerical analysis of the solutions was carried out using the \texttt{ode45} solver in \texttt{MATLAB}. This employs a Runge-Kutta algorithm, suitable for the numerical solution of non-stiff ordinary differential equations~\cite{Matlab}. The relative and absolute error tolerances were set to be $3\times10^{-14}$ and $3\times10^{-20}$ respectively. Numerical results were compared against analytic expectations (such as the Hubble rate being monotonically decreasing with time) and the Hamiltonian constraint was checked up to the set error tolerances.

\section{Two quadratic fields}

We now turn our attention to the case of two scalar fields, $\phi$ and $\psi$ with a potential given by:
\begin{equation}
V(\phi,\psi)=\frac{1}{2}m_{\phi}^2\phi^2+\frac{1}{2}m_{\psi}^2\psi^2
\end{equation}
 where $m_{\phi}$ and $m_{\psi}$ are the masses of the $\phi$ and $\psi$ fields respectively.

\begin{figure*}[h]
\centering
\begin{subfigure}[h]{0.6\textwidth}
\centering
\includegraphics[width=\textwidth]{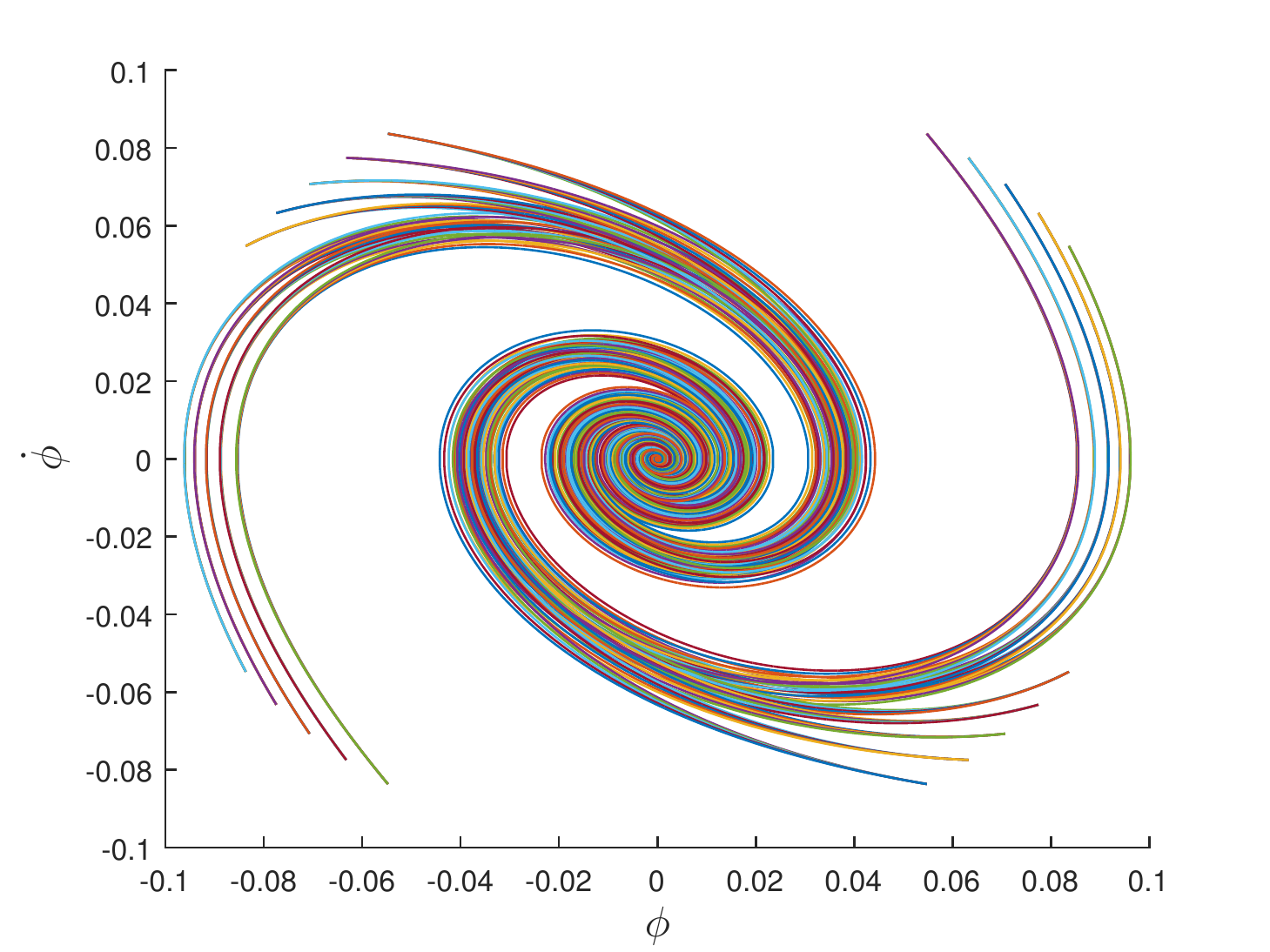}
\label{PhiPhidot}
\caption{$\phi$-$\dot{\phi}$ field projection.}
\end{subfigure}\\
\begin{subfigure}[h]{0.6\textwidth}
\centering
\includegraphics[width=\textwidth]{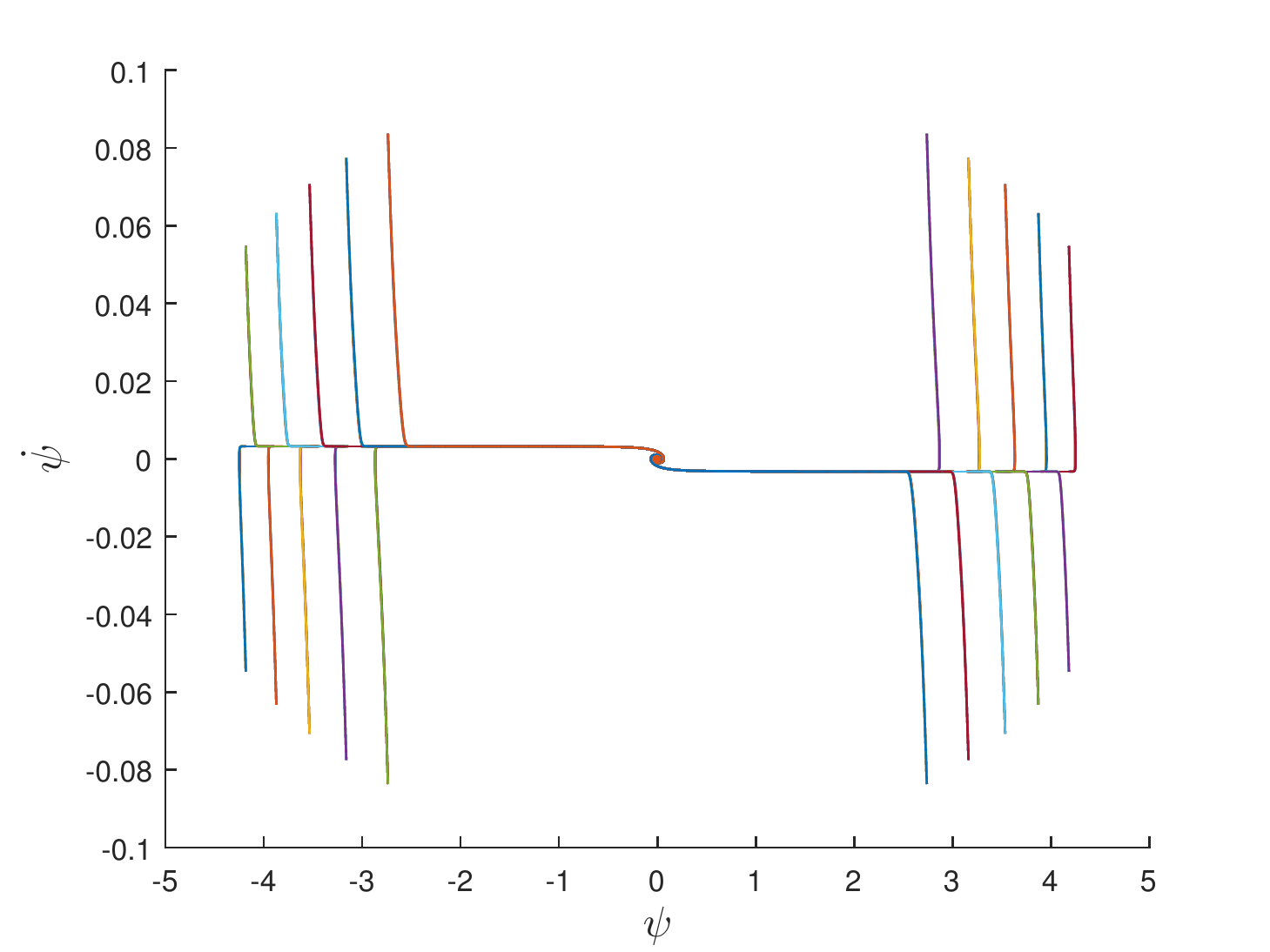}
\caption{$\psi$-$\dot{\psi}$ field projection.}
\label{PsiPsidot}
\end{subfigure}
\caption{The projections of the inflaton trajectories in the $\phi$-$\dot{\phi}$  and $\psi$-$\dot{\psi}$ planes for a selection of initial conditions, showing the attractor behaviour. We see the $\phi$ field spirals straight in towards the origin whereas the $\psi$ field first undergoes slow roll as it moves along one of the horizontal lines towards the origin. Masses used were $m_{\phi}=1$ and $m_{\psi}=0.02$, with an initial energy density $\rho_{0}=0.01$ in Planck units.}
\label{Phase}
\end{figure*}

\begin{figure}[h]
\centering
\includegraphics[width=0.6\textwidth]{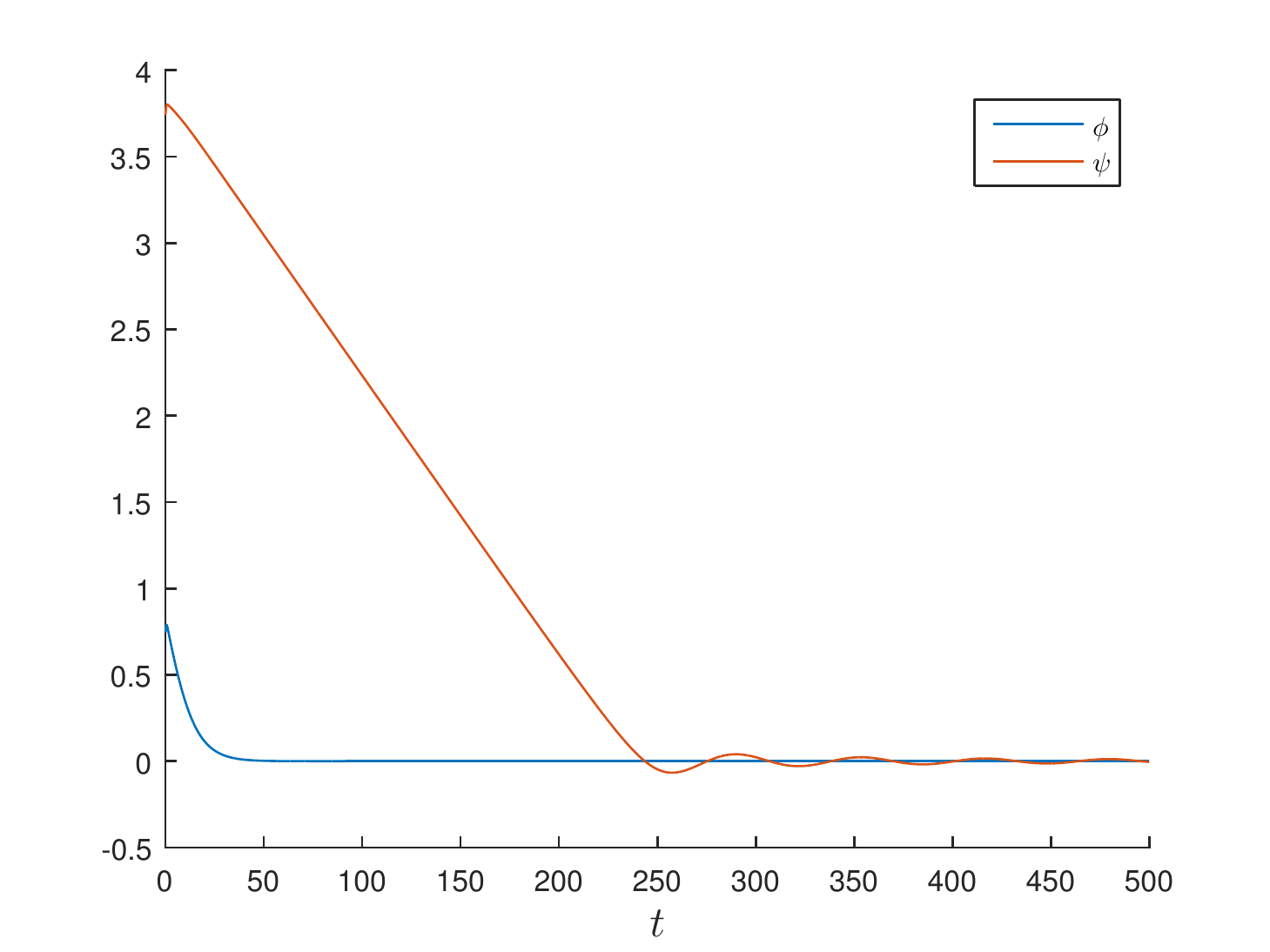}
\caption{The time evolution of the field values for a $\phi$ field of mass $m_{\phi}=0.5$ and a $\psi$ field of mass $m_{\psi}=0.1$, starting from an initial energy density of $\rho_{0}=0.2$ in Planck units. Here we can clearly see that, even with a factor of five difference, the heavier field reaches the potential minimum and exits the slow roll phase much faster than the lighter $\psi$ field.}
\label{TwoFieldEvol}
\end{figure}

Before we consider the issue of forming measures for this model, let us examine the dynamics of the scalar fields. We once again see inflationary attractors which, as explained in the introduction, are a result of the stretching of phase space trajectories along the volume axis. Figures~\ref{Phase} and~\ref{TwoFieldEvol} provide illustrative examples of the attractor behaviour and the light mass field dominance for the two quadratic fields.

\begin{figure}[h]
\centering
\includegraphics[width=0.6\textwidth]{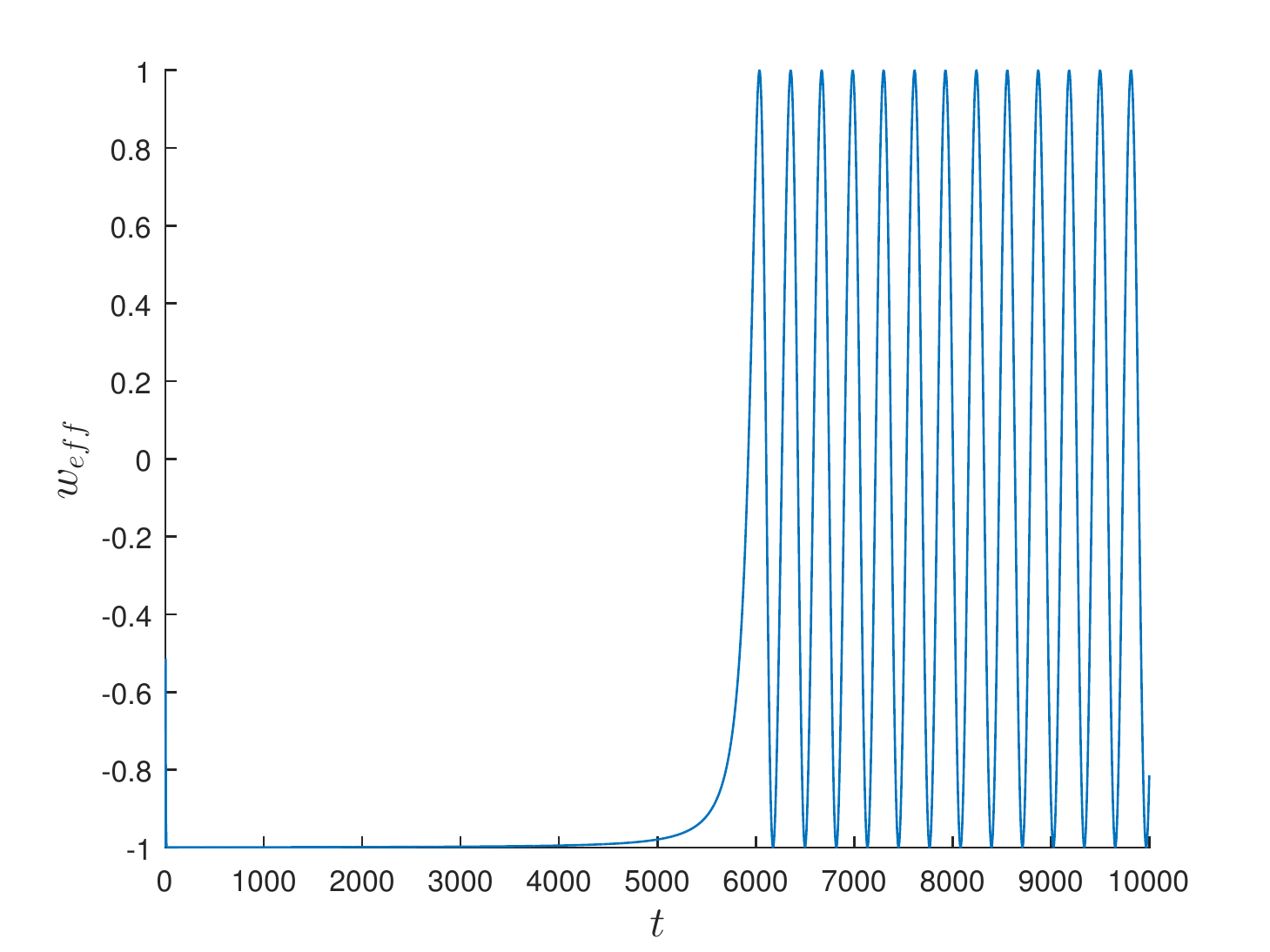}
\caption{The evolution of the effective equation of state parameter for the $\phi$ and $\psi$ fields, $w_{eff}$. Initially $w_{eff}$ falls towards $w_{eff}=-1$ as the dominant $\psi$ field undergoes slow roll, rising slowly until the inflation ends and $w_{eff}$ oscillates between -1 and 1.}
\label{w2}
\end{figure}

We note here that the dynamics of our system can be thought of as either a single inflaton rolling in the potential $V(\phi,\psi)$ subject to Hubble friction, or as two particles in one dimensional quadratic potentials. For figure~\ref{Phase} we set two masses differing by a factor of 50, which allows us to see two different kinds of field behaviour. In the case of the lighter field $\psi$ we see that it it has a larger region of phase space energetically available to it outside of the attractor. For the range of initial conditions the inflaton is very quickly driven towards slow roll (when the trajectories move along one of the horizontal lines) by the Hubble friction term, eventually passing into the attractor where the field oscillates about the potential minimum. For the heavier $\phi$ field the inflaton has far less phase space accessible to it. We observe no extended slow roll phase here, with the inflaton spiralling straight into the attractor. The behaviour of the effective equation of state parameter during inflation is shown in figure~\ref{w2}.

We now turn to looking at measures in this two field model. The phase space variables are $(v,H,\phi,P_{\phi},\psi,P_{\psi})$ so our symplectic structure is:
\begin{equation}
\omega=\dif v\wedge\dif H+\dif\phi\wedge\dif P_{\phi}+\dif\psi\wedge\dif P_{\psi}
\end{equation}
We wish to use this quantity form a measure on phase space. However, we must begin by asking ourselves where we should evaluate this measure in phase space. When counting trajectories in phase space we must be careful to count all trajectories exactly once. It so happens that in general relativity the Hubble parameter is monotonically non-increasing. To see this for $N$ fields in a given potential, we differentiate the Friedmann equation and use the scalar field equations of motion to obtain:
\begin{equation}
\dot{H}=-4\pi\sum_{i=1}^N\dot{\phi}_{i}^2
\end{equation}
It is clear that, given the fields are all real, $\dot{H}\leq0$. The case where $\dot{H}=0$ corresponds to all the field velocities being zero. These solutions correspond to all the matter existing in the form of a cosmological constant so are not of interest in general dynamical systems~\cite{SloanMin}. The fact that $H$ decreases monotonically with time and goes to infinity for each solution at the singularity means that each solution will cross any choice of constant $H$ surface exactly once. In order to ensure we only count trajectories once it is therefore convenient to form our measure on a surface of constant $H$, that is we will take a uniform distribution of initial conditions on some constant $H$ surface with respect to the symplectic measure.

Having established this, we now pull the symplectic structure back to a surface of constant $H$. On these surfaces $\dif H=0$ so our pullback is given by:
\begin{equation}
\underleftarrow{\omega}=\dif\phi\wedge\dif P_{\phi}+\dif\psi\wedge\dif P_{\psi}
\end{equation}
This can be expanded using $P_{\phi}=v\dot{\phi}$ and $P_{\psi}=v\dot{\psi}$ to give:
\begin{equation}
\underleftarrow{\omega}=v\dif\phi\wedge\dif\dot{\phi}+\dot{\phi}\dif{\phi}\wedge\dif v+v\dif\psi\wedge\dif\dot{\psi}+\dot{\psi}\dif{\psi}\wedge\dif v
\end{equation}
To obtain a measure that covers our phase space it must be raised to the power $2$, as mentioned in section 1.2. Using the antisymmetry of the wedge product we obtain:
\begin{equation}
\underleftarrow{\omega}^2=v\dot{\phi}\dif{\phi}\wedge\dif\psi\wedge\dif\dot{\psi}\wedge\dif v+v\dot{\psi}\dif\psi\wedge\dif\phi\wedge\dif\dot{\phi}\wedge\dif v
\end{equation}
When we wrote the expression for the Liouville measure in the introduction there were some normalization factors included. Probabilities are calculated by taking the ratio of measured sections meaning any normalization factors will cancel, therefore we shall not worry ourselves with them from here on.

Now that we have an expression for $\underleftarrow{\omega}^2$ we can in general calculate the probability of some event $X$ happening in our two field model by integrating the measure over the region of phase space where event $X$ occurs and dividing by the total measure, an integral over the whole of the phase space on our constant $H$ surface. For our purposes we shall focus on the simpler case of two equal mass fields. The integrals over the two separate terms will now be identical, given the symmetry of the problem. We therefore write the probability of event $X$ in this case as:
\begin{equation}
P(X)=\frac{\int_{S_{X}}v\dot{\psi}\dif\psi\dif\phi\dif\dot{\phi}\dif v}{\int_{S_{H}}v\dot{\psi}\dif\psi\dif\phi\dif\dot{\phi}\dif v}
\end{equation}
 where $S_{X}$ is the region over the constant $H$ hypersurface where event $X$ occurs and $S_{H}$ is the whole of the hypersurface.

Now we encounter the problem mentioned in the introduction. The integral over the volume direction is non-compact and yields infinity. This would seem to leave our probability meaningless. However as we saw from the Hamiltonian in equation~\ref{eq:13} our solutions remain physically identical under a rescaling in the volume $v\rightarrow\alpha v$, so we have a freedom in $v$ under re-scaling. The freedom comes about because $v$ is not a physical observable but rather a gauge choice. In order to talk about physical length scales in cosmology we must define some reference scale, the choice of which is arbitrary. A re-scaling of the volume factor $v$ simply corresponds to choosing a different reference scale and will yield a physically indistinguishable description of the system. Now, we do indeed obtain infinity by integrating over the whole of $v$ but in doing so we count the same physical solutions multiple times. Since we are interested only in the space of physically distinct solutions we can avoid the infinity by imposing a finite volume cut-off, $v_{\star}$, on the integrals in $P(X)$. The integrals over $v$ now cancel for any finite $v_{\star}$ yielding a well defined, cut-off independent probability on the space of physically distinct solutions. We also note here that the probability, defined in this way, only depends on physical degrees of freedom, the field values, field velocities and the Hubble parameter. This is to be expected, our calculations of probabilities should not have any dependence on gauge directions in our phase space~\cite{Sloan}.

Now we introduce the polar coordinates defined by $\phi=r\cos\theta$ and $\psi=r\sin\theta$. Changing to the variables $(\dot{\phi},r,\theta)$ and imposing the finite volume cut-off we obtain the following expression for the probability of event $X$ occurring in the case of two equal mass quadratic fields:
\begin{equation}
P(X)=\frac{1}{N_{H}}\int_{S_{X}}r\sqrt{\frac{3H^{2}}{4\pi}-m^{2}r^{2}-\dot{\phi}^{2}}\dif\dot{\phi}\dif r\dif\theta
\label{2P}
\end{equation}
 where $m$ is the mass of the two fields and $N_{H}$ is the total measure integrated over all the points consistent with the Hamiltonian and constant $H$ constraints i.e. the region bounded by:
\begin{equation}
m^{2}r^{2}+\dot{\phi}^{2}=\frac{3H^{2}}{4\pi}
\end{equation}

With all this at hand we are now in a position to calculate the probability that our model yields a universe observationally similar to our own. Precisely, starting at some initial Hubble surface $H_{i}$, what is the probability that our trajectories evolve to have a value of the spectral index, $n_{s}$, within the $68\%$ confidence interval reported by the Planck collaboration ($n_{s}=0.968\pm{0.006}$~\cite{Planck}) at some final Hubble surface $H_{f}$?

To calculate a numerical value for this probability we chose an initial Hubble value $H_{i}=9.15\times10^{-5}$ and a final Hubble value of $H_{f}=7.42\times10^{-6}$.\footnote{The value for $H_{i}$ was chosen here for primarily computational reasons, the value was roughly the largest value that would allow for a reasonable time-frame when numerically solving the equations of motion. The value for $H_f$ was chosen to correspond roughly with the Planck analysis. The exact value used will not have a huge effect on the probability calculation provided it is in the right region.} At the final Hubble surface we calculated the value of the of the spectral index using the expression in Bassett, Tsujikawa and Wands:
\begin{equation} 
n_{s}=1-6\epsilon+2\eta_{\sigma\sigma}
\end{equation}
where the slow roll parameters $\epsilon$ and $\eta_{\sigma\sigma}$ are defined as in section 1.1~\cite{Bassett}.

The masses used in the simulations were $m_{\phi}=m_{\psi}=1.1\times10^{-6}$. It was found that observational disagreement was obtained in these simulations when the inflaton started near the bottom of the potential, which is to be expected. By starting near the bottom of the inflationary potential the inflaton does not undergo the extended period of slow roll required to obtain an observationally consistent universe.

The exact region of observational disagreement in phase space is complicated in the multifield case by the fact that the inflaton now has an angular momentum. Depending on the directions and magnitudes of the initial field velocities the radius out to which we obtain observational disagreement will vary. If we start by pushing the inflaton up away from the minima then we will obtain more slow roll and vice versa. However, we can exploit this fact to obtain a lower bound on the probability. There will be some maximum radius $r_{0}$ where we can place the inflaton and find observational disagreement that corresponds to pushing the inflaton straight down to the minimum. If we take the value for $r_{0}$ and assume the region of observational agreement extends out to this radius within the constraint volume, the integral in equation \ref{2P} can be evaluated to give a lower bound on the probability of observational agreement. 

Given that the potential is symmetric in the equal mass case the procedure is to pick $\psi=\dot{\psi}=0$ and find the maximum value of $\phi$ at which we get observational disagreement, choosing $\dot{\phi}$ such that the inflaton is directed radially towards the potential minimum. Doing this we find that $r_{0}\approx3.839$. Evaluating the integral in equation \ref{2P} we obtain the following expression for the lower bound on the probability of observational agreement:
\begin{equation}
P_{L}(\mathcal{O})=1-\frac{m^{2}(\rho r_{0}^{2}-\frac{1}{4}m^{2}r_{0}^{4})}{\rho^{2}}
\end{equation}
 where $P_{L}(\mathcal{O})$ is the lower bound on the probability of observational agreement, $\mathcal{O}$ represents the event where we obtain observational agreement and $\rho$ is the initial energy density. For $r_{0}=3.839$ this gives a value of $P_{L}(\mathcal{O})=0.982$.

This lower bound already tells us that the probability of observational agreement is high for the chosen initial Hubble surface. As discussed in section 1.3, if we had taken an earlier Hubble surface then we would have obtained an even greater value for the probability. For comparison, if we carry out the single field calculation from the same initial Hubble surface we obtain $P(\mathcal{O})=0.9653$. The probability is greater in the two field case, a point that will be discussed in detail in section 6.

\section{Three quadratic fields}
Let us now consider the case of three quadratic fields so that the potential is given by:
\begin{equation}
V(\phi,\psi,\chi)=\frac{1}{2}m_{\phi}^{2}\phi^2+\frac{1}{2}m_{\psi}^{2}\psi^2+\frac{1}{2}m_{\chi}^{2}\chi^2
\end{equation}
 where $\phi$, $\psi$ and $\chi$ are the scalar fields and $m_{\phi}$, $m_{\psi}$ and $m_{\chi}$ the respective masses. 

The process of obtaining the measure and probability expressions is entirely analogous here to the two field case. The symplectic structure in the three field case is:
\begin{equation}
\omega=\dif v\wedge\dif H+\dif\phi\wedge\dif P_{\phi}+\dif\psi\wedge\dif P_{\psi}+\dif\chi\wedge\dif P_{\chi}
\end{equation}
 where $P_{\phi}$, $P_{\psi}$ and $P_{\chi}$ are the usual canonical momenta. As before, we take the pullback of this onto a surface of constant $H$ and raise it to the power $3$ in order to cover our constraint surface, giving:
\begin{equation}
\underleftarrow{\omega}^{3}=6v^{2}\dot{\psi}\dif\phi\wedge\dif\dot{\phi}\wedge\dif{\chi}\wedge\dif\dot{\chi}\wedge\dif\psi\wedge\dif v+6v^{2}\dot{\chi}\dif\phi\wedge\dif\dot{\phi}\wedge\dif{\psi}\wedge\dif\dot{\psi}\wedge\dif\chi\wedge\dif v+6v^{2}\dot{\phi}\dif\psi\wedge\dif\dot{\psi}\wedge\dif{\chi}\wedge\dif\dot{\chi}\wedge\dif\phi\wedge\dif v
\end{equation}
Restricting ourselves to the equal masses scenario once again, we can see that each of the integrals over the respective terms in $\underleftarrow{\omega}^{3}$ will be identical. We can therefore simply take the first term in $\underleftarrow{\omega}^{3}$ to form our probability.

Let us re-write the field values in terms of polar coordinates as $\phi=r\sin\theta\cos\xi$, $\psi=r\sin\theta\sin\xi$ and $\chi=r\cos\theta$. Imposing a finite volume cut-off as before we obtain the expression for the probability of event $X$ for three equal mass quadratic fields, given in equation \ref{3P}.
\begin{equation}
P(X)=\frac{1}{N_{H}}\int_{S_{X}}r^{2}\sin\theta\sqrt{\frac{3H^{2}}{4\pi}-m^{2}r^{2}-\dot{\phi}^2-\dot{\chi}^{2}}\dif\dot{\phi}\dif\dot{\chi}\dif r\dif\theta\dif\xi
\label{3P}
\end{equation}
Once again, $N_{H}$ is the total measure, this time integrated over the region bounded by the constraint hypersurface given by:
\begin{equation}
m^{2}r^{2}+\dot{\phi}^{2}+\dot{\chi}^2=\frac{3H^{2}}{4\pi}
\end{equation}
As in the two field case the exact form of the region of observational disagreement will depend on the size and magnitude of the initial field velocities. However we can obtain a lower bound by assuming the region of observational disagreement extends out to some radius $r_{0}$, where $r_{0}$ is found as in section 3. Integrating equation \ref{3P} we obtain the lower bound on the probability of observational agreement given in equation \ref{3PL}.
\begin{equation}
P_{L}(\mathcal{O})=1-\frac{1}{3\pi\rho^{3}}\left\{6\rho^{3}\arctan\left({\frac{mr_{0}}{\sqrt{2\rho-m^{2}r_{0}^{2}}}}\right)+mr_{0}\sqrt{2\rho-m^{2}r_{0}^2}\left(7m^{2}\rho r_{0}^{2}-2m^{4}r_{0}^{4}-3\rho^{2}\right)\right\}
\label{3PL}
\end{equation}
Using the same $H_{i}$, $H_{f}$ and $m$ as before we find $r_{0}=3.839$ and $P_{L}(\mathcal{O})=0.997$ for this model. We note again that the probability of observational agreement is found to increase with the number of fields.

\section{An $N$-field measure}

Let us now turn to consider what the measure will be for a general $N$-field model given by the Hamiltonian in equation \ref{eq:13}. The symplectic structure is given by:

\begin{equation}
\omega=\dif v\wedge\dif H+\sum_{i=1}^{N}\dif\phi_{i}\wedge\dif P_{i}
\end{equation}

Taking the pullback of this onto a surface of constant Hubble rate and raising it to the power $N$ will result in two sets of terms. The first will involve the wedge products of the $v\dif\phi_{i}\wedge\dif\dot{\phi}_{i}$ terms with  one another. The second will consist of the product of the $\dot{\phi}_{i}\dif\phi_{i}\wedge\dif v$ terms with all the $v\dif\phi_{j}\wedge\dif\dot{\phi}_{j}$ terms where $j\neq i$. All other terms in the expansion will contain a wedge product sequence with identical differentials and will therefore vanish due to the antisymmetry of the wedge product. Expanding out $\underleftarrow{\omega}^{N}$ we obtain equation \ref{N1}.
\begin{equation}
\underleftarrow{\omega}^{N}=\sum_{P}\prod_{j}(v\dif\phi_{j}\wedge\dif\dot{\phi}_{j})+Nv^{N-1}\sum_{k}\dot{\phi}_{k}\dif\phi_{k}\wedge\dif v\wedge\left\{\sum_{P}\prod_{j\neq k}(\dif\phi_{j}\wedge\dif\dot{\phi}_{j})\right\}
\label{N1}
\end{equation}
where $\sum_{P}$ is instructing us to sum over the permutations of the indices. All the $\dif\phi_{k}$ terms have been moved to the left in the second term by exploiting the fact that moving a pair of differentials together through the wedge product sequence will always return a positive sign.

Now, on a surface of constant $H$ all the configurations of $\prod_{j}(v\dif\phi_{j}\wedge\dif\dot{\phi}_{j})$ are equal to zero. This can be seen by writing the differential $\dif\phi_{j}$ in terms of the other field differentials. This will give a collections of terms with two identical terms in the wedge product sequence which will therefore vanish. Finally we note that we can just pick one configuration of the $\prod_{j\neq k}(\dif\phi_{j}\wedge\dif\dot{\phi}_{j})$ terms, bringing out a factor of $(N-1)!$ from the permutations. This gives us the final expression for the $N$-field measure:
\begin{equation}
\underleftarrow{\omega}^{N}=N!v^{N-1}\sum_{k}\dot{\phi}_{k}\dif\phi_{k}\wedge\dif v\wedge\prod_{j\neq k}(\dif\phi_{j}\wedge\dif\dot{\phi}_{j})
\end{equation}
This expression was checked explicitly against the two and three field results derived previously and holds for a general $N$-field model. To implement it in probability calculations, the procedure is the same as in the two and three field cases, one should impose a finite volume cut-off and integrate over the appropriate regions of phase space.

\section{The limit as $N\rightarrow\infty$}

It has been shown that as we increase the number of fields in the equal mass quadratic model from one to three, the probability of observational agreement increases.. We now consider the probability as the number of fields becomes very large.

In order to get a hold of the behaviour as we increase the number of fields we consider the surface of revolution for the potential $V=m^{2}r^{2}/2$, where $r=\sum_{i}\phi_{i}^{2}$. When selecting initial conditions we place the inflaton somewhere on the potential surface. The probability of observational disagreement will scale with the ratio of the surface of revolution for the region of observational disagreement to the total surface of revolution for the potential.

We now assume the region of observational disagreement extends to some radius $r_{0}$. The total allowed potential surface extends out to the radius $R_{H}=\sqrt{3H_{i}^{2}/4\pi m^{2}}$, which is the maximum allowed radius given an initial Hubble rate, $H_{i}$. We now wish to calculate the surface of revolution for the potential function $V$. To obtain the $N$-dimensional surface of revolution for our $N$-field potential function we must rotate it through the surface of an $(N-1)$-sphere. The surface area for an $(N-1)$-sphere of radius $r$ is given by:
\begin{equation}
S_{N-1}=\frac{2\pi^{N/2}}{\Gamma\left(N/2\right)}r^{N-1}
\end{equation}  
where $\Gamma(x)$ is the Gamma function~\cite{Wolfram}. 

The surface of revolution for our region of observational disagreement is given in the usual way by:
\begin{equation}
S_{rev}(r_{0})=\frac{2\pi^{N/2}}{\Gamma\left(N/2\right)}\int_{0}^{r_{0}}r^{N-1}\sqrt{1+\left(\frac{\dif V}{\dif r}\right)^{2}}\dif r
\end{equation}
Evaluating this integral we obtain:
\begin{equation}
S_{rev}(r_{0})=\frac{2\pi^{N/2}r_{0}^{N}}{N\Gamma\left(N/2\right)}\,_{2}F_{1}\left(-\frac{1}{2},\frac{N}{2},\frac{N}{2}+1,-m^{4}r_{0}^{2}\right)
\end{equation}
where $\,_{2}F_{1}\left(-\frac{1}{2},\frac{N}{2},\frac{N}{2}+1,-m^{4}r_{0}^{2}\right)$ is a hypergeometric function. Evaluating the expression in the same manner for the whole surface of revolution, we obtain the following expression for the ratio of the surfaces of revolution:
\begin{equation}
\zeta=\left(\frac{r_{0}}{R_{H}}\right)^{N}\frac{\,_{2}F_{1}\left(-\frac{1}{2},\frac{N}{2},\frac{N}{2}+1,-m^{4}r_{0}^{2}\right)}{\,_{2}F_{1}\left(-\frac{1}{2},\frac{N}{2},\frac{N}{2}+1,-m^{4}R_{H}^{2}\right)}
\end{equation}
The hypergeometric functions in this expression are both $\sim1$ for relevant values of $m^{4}r_{0}^{2}$ and $m^{4}R_{H}^{2}$. Given that $r_{0}<R_{H}$, it can be seen that as $N\rightarrow\infty$ we have $\zeta\rightarrow0$.

From this we conclude that as $N\rightarrow\infty$ the probability of observational agreement approaches one. The increase in the probability of observational agreement found in sections 3 and 4 is to be expected from the scaling of surface areas as we increase the dimension of phase space.

This result holds independent of our choice of $H_{i}$. The only requirement is that the region of observational disagreement is a subset of our total phase space, which we take to be a priori true. This means that provided we have a sufficient number of fields we can obtain a high probability of observational agreement, even if we choose to start from a late Hubble surface as in Gibbons and Turok~\cite{Turok}.

\section{Conclusions}

Let us now summarise what has been found. We have derived expressions for the two and three field measures and used these to obtain explicit expressions for the probabilities of observational agreement in the equal mass scenario. Starting from the initial Hubble surface $H_{i}=9.15\times10^{-5}$ and assuming the region of observational disagreement extended out to the radius $r_{0}$, we calculated lower bounds on the probabilities of observational agreement. For the equal mass two and three field models these were found to be $0.982$ and $0.997$ respectively, both of which were greater than the single field probability of $0.9653$.

It should be noted here that there is an important ambiguity in these probability calculations, that is the choice of initial Hubble surface. In principle, we could have chosen to form the measure on an earlier or later Hubble surface and would have obtained a different value for the probability of observational agreement. This issue remains unresolved, the key controversy being over whether to choose a uniform distribution of initial conditions early or late in the inflation.

We presented a derivation of the symplectic measure for a general $N$-field inflation model. This can be implemented in any future work by following the standard procedure of imposing a finite volume cut-off and integrating over the appropriate regions of phase space. This measure can be used in a detailed numerical study with \texttt{MULTIMODECODE} to precisely determine the regions of observational disagreement. 

Finally the increase in probability with the number of fields was found to be a result of the scaling of surface areas as we increase the number of fields and hence the dimension of our phase space. Further it was argued that as the number of fields approaches infinity the probability of observational agreement approaches one. This result was robust under a different choice of initial Hubble surface meaning that, provided we have a sufficient number of fields, we can obtain a high probability of observational agreement irrespective of our choice of initial Hubble surface.

Further extensions of this work include studying models with a range of field masses, introducing field interactions into the potential, and going beyond the FRW case to consider anisotropic Bianchi cosmologies. Here we have presented analytic results obtained in the case of symmetries, which indicate that increasing the number of fields increases the likelihood of agreement with observations (given appropriate inflaton masses). Further we have provided the theoretical framework from which a more rigorous exploration of the likelihoods can be analysed in the more complex multifield approach. When different field masses are used we expect the lightest fields to dominate the slow roll phase, with the heavier fields reaching the oscillatory phase very quickly in comparison. This behaviour should be apparent in any numerical exploration of different mass models. Moreover, we expect the argument presented in section VI can be extended here to include different masses. Provided the heaviest field starts above a certain point on the potential surface we expect there will be enough inflation for observational agreement. Extending the same scaling arguments we should see the probability of observational agreement approach one as the number of fields approaches infinity in the different mass case as well.

\section{Acknowledgements}

The authors thank Prof. Roger Davies for his invaluable comments and advice throughout this work. DS acknowledges support from a grant from the John Templeton Foundation.

\end{document}